\documentclass[superscriptaddress,aps,pra,twocolumn,showpacs,floatfix]{revtex4-2}
\usepackage[utf8]{inputenc}
\usepackage{graphicx,amsmath,amsfonts,amssymb}
\usepackage{color}
\usepackage[colorlinks=true, allcolors={blue}]{hyperref}
\usepackage{graphicx}
\usepackage{epstopdf}
\usepackage{float}
\usepackage{placeins}
\newcommand{\orcid}[1]{\href{https://orcid.org/#1}{\includegraphics[width=7pt]{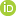}}}

\begin{document}
\preprint{APS/123-QED}

\title{Enhanced frequency estimation by non-Gaussianity of Fock states}

\author{Jonas F. G. Santos \orcid{0000-0001-9377-6526}}
\email{jonassantos@ufgd.edu.br}
\affiliation{Faculdade de Ci\^{e}ncias Exatas e Tecnologia, Universidade Federal da Grande Dourados, Caixa Postal 364, CEP 79804-970, Dourados, MS, Brazil}

\begin{abstract}
Leveraging the unique quantum properties of non-Gaussian states is crucial for advancing continuous variable quantum technologies. Recent experimental advancements in generating non-Gaussian states, coupled with theoretical findings of their superior performance in quantum information protocols compared to Gaussian states, motivate this investigation. This work investigates the impact of non-Gaussianity on the precision of frequency estimation using a quantum probe. We analyze a single bosonic mode and its non-Gaussian excited states as a system, while the frequency estimation
is investigated by explicitly computing the quantum Fisher information.
Our results demonstrate a significant enhancement in the quantum Fisher information for non-Gaussian states compared to Gaussian states with equivalent second-order moments. Importantly, we find that the increased quantum Fisher information achieved with non-Gaussian states outweighs their higher energetic cost compared to the Gaussian ground state.
We also briefly discuss two photonic non-Gaussian state generation schemes: interaction with an ancillary system and the use of photon number measurement on the photonic probe. 
\end{abstract}

\maketitle

\section{Introduction}

Quantum information processing based on continuous variable (CV) systems
has been widely investigated in the last years due to its mathematical
framework and the ability to be implemented in different experimental
platforms. Quantum metrology and sensing protocols have attracted
considerable attention in this scenario. The main purpose of quantum
metrology is to search for maximum precision limits in estimating
relevant parameters by exploiting intrinsic quantum features of probes,
such as coherence and quantum correlations \citep{Santos2025,Lee2020,Porto2025}.
The current state-of-the-art in quantum metrology and sensing includes
gravitational waves detectors \citep{Schnabel2010,Danilishin2020},
measurements in gravimetry and magnetometry \citep{Hou2020,Qvarfort2018},
and the investigation of metrological power with non-Hermitian systems
\citep{Wang2024,Ding2023,Santos2025A}. In addition to theoretical developments,
experimental progress has been reported using trapped-ions systems
\citep{Marciniak2022}, superconducting systems \citep{Wolski2020},
and single photons \citep{Pirandola2018}. In CV systems, most natural
probe states are those described by Gaussian states, for example,
the vacuum and thermal states \citep{Weedbrook2012,Serafini2017, Santos2025B}.
They have been used to estimate physical parameters such as frequency
and temperature \citep{Binder2020,Cenni2022} as well as parameters
associated with Gaussian unitary channels \citep{Pinel2013}. 

Despite the advance in quantum metrology based on Gaussian states,
the possibility of exploring non-Gaussian states for different quantum protocols has emerged in recent years \cite{Lvovsky2020, Rácz2024, Rakhubovsky2025, Wang2017, Wei2025, Zhang2024A}. In particular for quantum metrology and sensing, the use of non-Gaussian states as quantum probes has also progressed as a new route to access other intrinsic
quantum features \citep{Hanamura2021,Deng2024,Oh2020,Perarnau-Llobet2020,Tatsuta2019,Stammer2024,Rahman2025, Chen2023}.
In fact, Ref. \citep{Albarelli2018} introduced a resource theory
for non-Gaussianity, in which it is shown that a suitable measure of
non-Gaussianity based on the relative entropy, first presented in
Ref. \citep{Genoni2010}, fulfills all properties to be a bona-fide
measure to quantify the non-Gaussian distribution in a quantum state.
Non-Gaussian states have been employed in estimation of random displacement
\citep{Hanamura2021,Deng2024} as well as in phase estimation in interferometric
scheme \citep{Perarnau-Llobet2020,Deng2024,Oh2020}. Furthermore,
metrological tasks have been developed using cat states \citep{Tatsuta2019,Stammer2024}.
These theoretical advances are consistent with experimental techniques
to generate non-Gaussian states using, for instance, a superconducting
microwave cavity \citep{Deng2024} and a mechanical resonator \citep{Rahman2025}.
These recent efforts demonstrate that non-Gaussianity is a powerful
intrinsic quantum feature that can be exploited to characterize quantum
states, i.e., by estimating the temperature or frequency, and even
to estimating parameters of relevant channels. Regarding the frequency
estimation, different approaches have been developed recently, including
the use of Gaussian operations in a quantum harmonic oscillator \citep{Binder2020},
the phase transition in the Rabi model \citep{Garbe2020}, and non-linearity
in quantum systems \citep{Montenegro2025}. 

Although some important contributions have shown that non-Gaussian
features can indeed be useful to improve parameter estimation, the
direct relation between the improvement in the estimation and a measure
of non-Gaussianity degree is unclear so far. In this work, we aim
to shed light in this direction, by investigating the frequency estimation
using a single photonic mode as a probe system, while the excited
states play the role of probe states. For this purpose, we compute
the quantum Fisher information (QFI) for the frequency estimation
and explore the non-Gaussian feature of excited states of the photonic
mode to enhance the QFI. The importance of the QFI is that it is related
to the Cramér-Rao bound \citep{Safranek2015}, which sets a fundamental
lower limit on the mean squared error of a parameter to be estimated.
Thus, a higher QFI may imply a reduction in the uncertainty of the
frequency. At the same time, we provide an explicit expression for
the non-Gaussianity degree based on the relative entropy for the probe
states. The direct connection between the QFI and the non-Gaussianity
degree is also discussed, a result that could be experimentally tested
by measuring the covariance matrix for each excited state through
the reconstruction of the Wigner function.

The present work is organized as follows. Section \ref{sec:Quantifying-non-Gaussianity-of}
is dedicated to introduce the relevant tools to deal with continuous
variable systems, as the statistical moments. We also quantify explicitly
the non-Gaussianity degree for the excited states of a photonic mode.
In section \ref{sec:Frequency-estimation-with} we present the main
results concerning the frequency estimation, computing the QFI in
terms of the excitation mode and comparing with the Gaussian analog
result. We discuss the energetic cost of using non-Gaussian states
by relating the QFI to the energy difference between an excited state
and the corresponding ground state, the latter having a Gaussian distribution.
The evolution of the QFI as a function of the non-Gaussianity degree
for each excited state as probe is also made. Additionally, we briefly
discuss in section \ref{sec:Generating-nG-degree-in} two possibilities
of generating non-Gaussian states in a photonic probe, i.e., transferring
non-Gaussianity from an ancillary photonic mode or by applying photon
number measurements and having sufficient control over the measurement
parameter. In section \ref{QFIsuperpositionsection} is discussed the QFI for a Fock-state superposition. Finally, the conclusion and final remarks are discussed
in section \ref{sec:Conclusion}.

\section{Quantifying non-Gaussianity of Fock states\label{sec:Quantifying-non-Gaussianity-of}}

In order to introduce non-Gaussianity of Fock states, we firstly review
the main properties defining the class of continuous-variables (CV)
states. We here consider a single photonic mode described by the annihilation
and creation field operators $a$($a^{\dagger}$), but the formalism
can be easily extended for $N$ bosonic modes \citep{Braunstein2005,Weedbrook2012,Serafini2017}.
The field operators can be used to write the quadrature operators
for position and momentum, $q=\sqrt{\hbar/\left(2\omega\right)}\left(a^{\dagger}+a\right)$
and $p=i\sqrt{\hbar\omega/2}\left(a^{\dagger}-a\right)$, with $\omega$
the characteristic transition frequency between the different energy
levels. Any quantum state $\rho$ describing this photonic system
can be represented by its characteristic function 

\begin{equation}
\chi_{\rho}\left(\alpha\right)=Tr\left[\rho D\left(\alpha\right)\right],
\end{equation}
with $D\left(\alpha\right)=\exp\left[\alpha a^{\dagger}-\alpha^{\ast}a\right]$
the single-mode displacement operator and $\alpha\in\mathbb{C}$.
The characteristic function allows an alternative way to describe
quantum states, i.e., in terms of the Wigner function, which is the
Fourier transform of the characteristic function,

\begin{equation}
W_{\rho}\left(\lambda\right)=\frac{1}{\pi^{2}}\int d\alpha e^{\alpha^{\ast}\lambda+\lambda^{\ast}\alpha}\chi_{\rho}\left(\alpha\right).
\end{equation}

By introducing the vector of quadrature operators $R=\left(q,p\right)^{T}$,
any quantum state $\rho$ can be also fully described by its statistical
moments. The first moments are simply the mean values and are given
by $d_{i}=\langle R_{i}\rangle_{\rho}$. In addition, the second moments
are relevant to characterize uncertainty and squeezing properties
and are defined by the matrix elements $\sigma_{ij}=\langle R_{i}R_{j}+R_{j}R_{i}\rangle_{\rho}-2\langle R_{i}\rangle_{\rho}\langle R_{j}\rangle_{\rho}$.
The third moments (skewness) are given by $Tr\left[\rho\left(R_{i}-\langle R_{i}\rangle_{\rho}\right)^{3}\right]$,
while the fourth moments (kurtosis) are given by $Tr\left[\rho\left(R_{i}-\langle R_{i}\rangle_{\rho}\right)^{4}\right]$.
All these moments are associated with the distribution in the phase-space
and can be visualized using the Wigner function representation for
a quantum state. A special class of states describing CV systems is
that formed by Gaussian states. A Gaussian state $\rho_{G}$ is defined
by its Wigner function, which has a Gaussian distribution, and is
completely characterized by the first and second moments, with the
moments of superior order being zero. On the other hand, for non-Gaussian states the higher moments must be considered. Despite the possibility
of calculating all moments associated to a non-Gaussian state, they
do not provide an easy way to be employed in a resource theory. The
interest in using non-Gaussian states in quantum tasks has grown considerably
in the last years, together with the ability to generate these states
in different experimental platforms. In order to quantify the non-Gaussianity (nG)
degree of an arbitrary state $\rho$, Ref. \citep{Genoni2010} proposed
a suitable quantity based on the quantum relative entropy, defined
by $S\left[\rho_{1}||\rho_{2}\right]=Tr\left[\rho_{1}\left(\ln\rho_{1}-\ln\rho_{2}\right)\right]$,
with $\rho_{1}$ and $\rho_{2}$ two general quantum states. The non-Gaussianity
degree (nG-degree) is introduced as being $\delta_{nG}\left[\rho\right]=S\left[\rho||\rho_{G}\right]$,
where $\rho_{G}$ is a reference state which has the same first and
second statistical moments of $\rho$. The quantity $\delta_{nG}\left[\rho\right]$
can be written as \citep{Genoni2010}

\begin{equation}
\delta_{nG}\left[\rho\right]=S\left(\rho_{G}\right)-S\left(\rho\right),\label{ng01}
\end{equation}
with $S\left(\rho\right)$ the von Neumann entropy. The quantity $\delta_{nG}\left[\rho\right]$
fulfills some relevant properties, for instance, it is invariant under
unitary transformation $U$, $\delta_{nG}\left[U\rho U^{\dagger}\right]=\delta_{nG}\left[\rho\right]$,
as well as $\delta_{nG}\left[\rho\right]=0$ iff $\rho$ is a Gaussian
state. In particular, the first property implies that the nG-degree does not change under squeezing, rotation, and displacement operations.

Let us consider a photonic system described by a single quantum harmonic
oscillator with frequency $\omega$ as the standard model which, in
the next section, will play the role of a probe system. The Hamiltonian
is given by $H=\hbar\omega a^{\dagger}a$, with $a^{\dagger}$($a$)
standing for the creation (annihilation) bosonic operators. The eigenstates
in the Fock basis $\left\{ |n\rangle\right\} _{n=0}^{\infty}$ are
simply $|n\rangle$ with energies $E_{n}=\hbar\omega n$. It is well-known
that only the ground state $|0\rangle$ has a Gaussian distribution.
Written the Fock states in the position representation $\langle x|n\rangle=|\psi_{n}\left(x\right)\rangle$,
one has

\begin{equation}
|\psi_{n}\left(x\right)\rangle=\left(\frac{\omega}{\pi}\right)^{1/4}\frac{1}{\sqrt{2^{n}n!}}He_{n}\left(\sqrt{\omega}x\right)\exp\left[-\omega x^{2}/2\right],\label{psi01}
\end{equation}
where $He_{n}\left(\sqrt{\omega}x\right)$ is the Hermite polynomial
of degree $n$, with $H_{0}\left(\sqrt{\omega}x\right)=1$, resulting
in the Gaussian distribution for the ground state. Here, we are interested
in the non-Gaussianity degree $\delta_{nG}\left[\rho\right]$ for
the excited Fock states, which will be the probe states in the next
section. Written the density operator for each Fock state as $\rho_{n}=|\psi_{n}\left(x\right)\rangle\langle\psi_{n}\left(x\right)|$
and noting that $\rho_{n}$ is a pure state for every excitation $n$,
the quantity $\delta_{nG}\left[\rho_{n}\right]$ is explicitly given
by

\begin{equation}
\delta_{nG}\left[\rho_{n}\right]=\left(n+1\right)\ln\left(n+1\right)-n\ln n.\label{ng02}
\end{equation}

\begin{figure}
\includegraphics[scale=0.4]{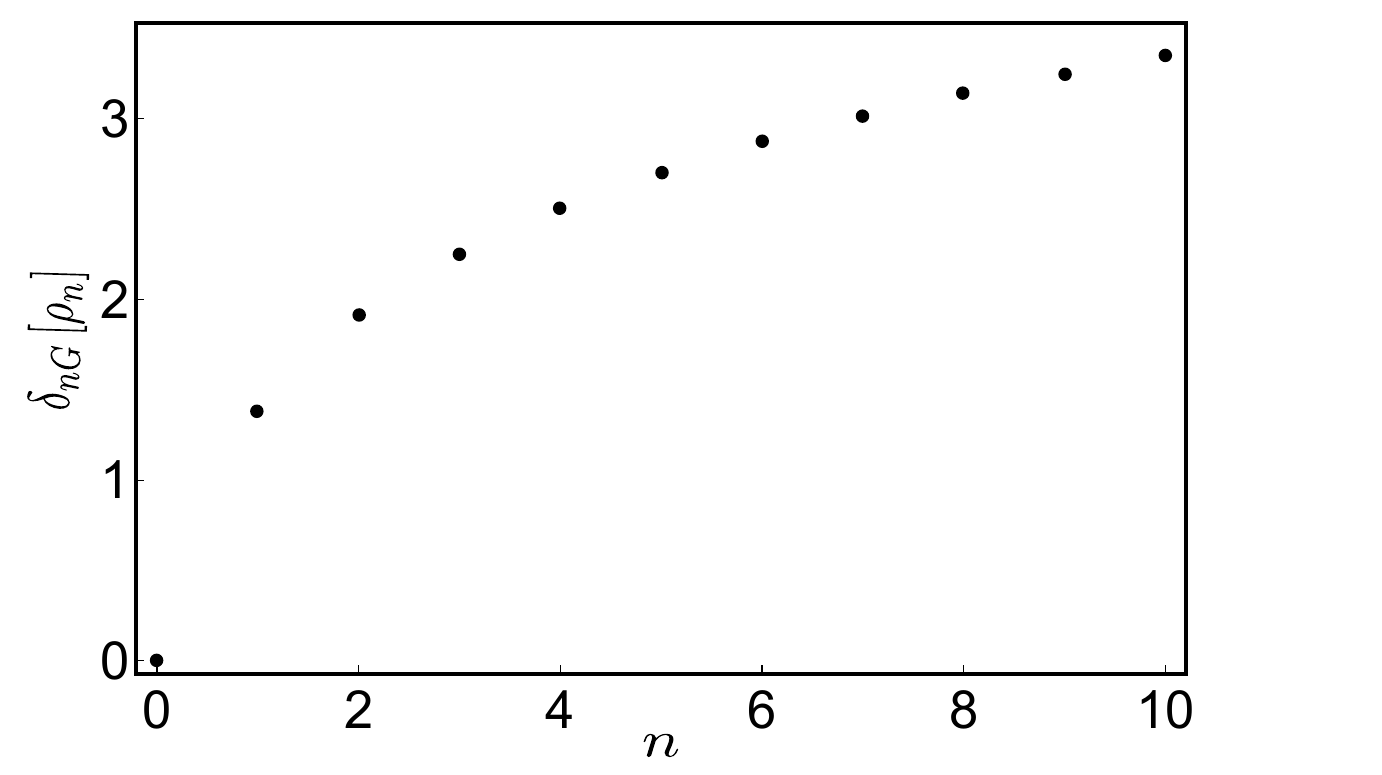}

\caption{Non-Gaussianity quantifier for the Fock states of the quantum harmonic
oscillator, $\delta_{nG}\left[\rho_{n}\right]$, for the first ten
pure states.}

\label{nGfigure}
\end{figure}

Note that, for the ground state, $n=0$ and then $\delta_{nG}\left[\rho_{0}\right]=0$,
since $\text{limit}_{n\rightarrow0}\left(n\ln n\right)=0$. Equation
(\ref{ng02}) shows that the more excited the Fock state, the higher
the non-Gaussianity degree. Figure \ref{nGfigure} depicts the nG-degree
$\delta_{nG}\left[\rho_{n}\right]$ as a function of $n$ for the
first 10 excited Fock states of a single quantum harmonic oscillator. It must be stressed that $\delta_{nG}\left[\rho_{n}\right]$ depends only on the photon number $n$ and is independent of the frequency.

\section{Frequency estimation with Fock states\label{sec:Frequency-estimation-with}}

Fock states of a single quantum harmonic oscillator are examples of
simple pure states that can be generated using different techniques
\citep{Lvovsky2020}. Pure states do not carry coherence or entanglement,
two quantum features often exploited in several parameter estimation
protocols. Instead, they carry non-Gaussianity which is quantified
by Eq. (\ref{ng02}), and a resource theory for non-Gaussianity has
been established in Ref. \citep{Albarelli2018}. The use of non-Gaussian
states in quantum metrology protocols has been considered recently
in some scenarios \citep{Hanamura2021,Deng2024,Oh2020,Perarnau-Llobet2020,Tatsuta2019,Stammer2024,Rahman2025},
but its connection and quantification in terms of the degree of non-Gaussianity
for each probe state remains an open question. 

We here consider the frequency estimation of a single quantum harmonic
oscillator employing pure Fock states $\rho_{n}\left(\omega\right)$
as probes and the use of the non-Gaussianity to enhance the parameter
estimation. To do so, we denote the squared sensitivity for the estimation
of the frequency $\omega$ by $\left(\delta\omega\right)^{2}$. After
$N$ measurements of an arbitrary operator $A$ with measurements
results $a_{i}$, the precision in estimating $\omega$ is bounded
from below by the Cramér-Rao bound

\begin{equation}
\left(\delta\omega\right)^{2}\geq\frac{1}{N\mathcal{F}_{\omega}},\label{qfi01}
\end{equation}
where $\mathcal{F}_{\omega}=\mathcal{F}_{\omega}\left[\rho_{n}\left(\omega\right)\right]$
is the quantum Fisher information (QFI), obtained by the optimization
of the classical Fisher information over all possible positive operator-valued
measure (POVM) \citep{Braunstein1994}. For pure states $\rho_{n}=|\psi_{n}\left(x\right)\rangle\langle\psi_{n}\left(x\right)|$
the QFI is given by

\begin{equation}
\mathcal{F}_{\omega}=4\left(\langle\partial_{\omega}\psi_{n}\left(x\right)|\partial_{\omega}\psi_{n}\left(x\right)\rangle-|\langle\psi_{n}\left(x\right)|\partial_{\omega}\psi_{n}\left(x\right)\rangle|^{2}\right).\label{qfi02}
\end{equation}

Equations (\ref{qfi01}) and (\ref{qfi02}) imply that the more sensitive
the probe state is to small variations of the frequency, the higher
the QFI and consequently the precision in estimating $\omega$ can
be enhanced. From Eq. (\ref{psi01}) the quantities on the right side
of Eq. (\ref{qfi02}) can be analytically calculated by noting that 

\begin{align}
\langle\partial_{\omega}\psi_{n}\left(x\right)|\partial_{\omega}\psi_{n}\left(x\right)\rangle & =\int_{-\infty}^{\infty}\dot{\psi}_{n}^{\ast}\left(x\right)\dot{\psi}_{n}\left(x\right)dx,\\
\langle\psi_{n}\left(x\right)|\partial_{\omega}\psi_{n}\left(x\right)\rangle & =\int_{-\infty}^{\infty}\psi_{n}^{\ast}\left(x\right)\dot{\psi}_{n}\left(x\right)dx,
\end{align}
with dots representing derivative with respect to the frequency, which
leads to the following expression for the QFI

\begin{equation}
\mathcal{F}_{\omega}\left[\rho_{n}\left(\omega\right)\right]=\frac{n^{2}+n+1}{2\omega^{2}}.\label{qfi03}
\end{equation}

The QFI in Eq. (\ref{qfi03}) is similar to the result derived in
Ref. \citep{Volkoff2025}. However, Ref. \citep{Volkoff2025} does
not discuss the source of possible enhancement in the estimation of
the parameter studied. Given that Gaussian states are employed in
many quantum protocols and are sometimes called the most classical
quantum state, it is interesting to obtain the QFI for Gaussian states
$\rho_{n}^{G}\left(\vec{d},\boldsymbol{\sigma}\right)$, with the
same first moments and covariance matrix as the Fock state $\rho_{n}$,
i.e., $\vec{d}=\langle d_{i}\rangle_{\rho_{n}}$ and $\sigma_{ij}=\langle d_{i}d_{j}+d_{j}d_{i}\rangle_{\rho_{n}}-2\langle d_{i}\rangle_{\rho_{n}}\langle d_{j}\rangle_{\rho_{n}}$.

For a single-mode Gaussian state $\rho^G_\theta$, the QFI depends only on the first and second moments and is given by
\begin{equation}
\mathcal{F}\left(\rho_{\theta}^G\right)=\frac{1}{2}\frac{\text{Tr}\left[\left(\boldsymbol{\sigma}_{\theta}^{-1}\boldsymbol{\sigma}_{\theta}'\right)^{2}\right]}{1+P_{\theta}^{2}}+2\frac{(P_{\theta}^{'})^{2}}{1-P_{\theta}^{4}}+\Delta\vec{d}'^{T}\boldsymbol{\sigma}_{\theta}^{-1}\Delta\vec{d}',
     \label{FisherGaussian}
\end{equation}
where $P_\theta = |\boldsymbol{\sigma}_\theta|^{-2}$ represents the purity of the one-mode Gaussian state, $\Sigma_{\theta}^{-1}$  is the inverse matrix of $\Sigma_\theta$, $\boldsymbol{\sigma}_\theta'$ and $P_{\theta}^{'}$ denotes the differentiation of the covariance matrix and purity concerning the parameter $\theta$, respectively. Applying Eq. (\ref{FisherGaussian}) to the Gaussian state $\rho_{n}^{G}\left(\vec{d},\boldsymbol{\sigma}\right)$ we obtain
\begin{equation}
\mathcal{F}_{\omega}\left[\rho_{n}^{G}\left(\vec{d},\boldsymbol{\sigma}\right)\right]=\frac{\left(2n+1\right)^{2}}{\left[\left(2n+1\right)^{2}+1\right]\omega^{2}}.\label{qfi04}
\end{equation}

\begin{figure}
\includegraphics[scale=0.36]{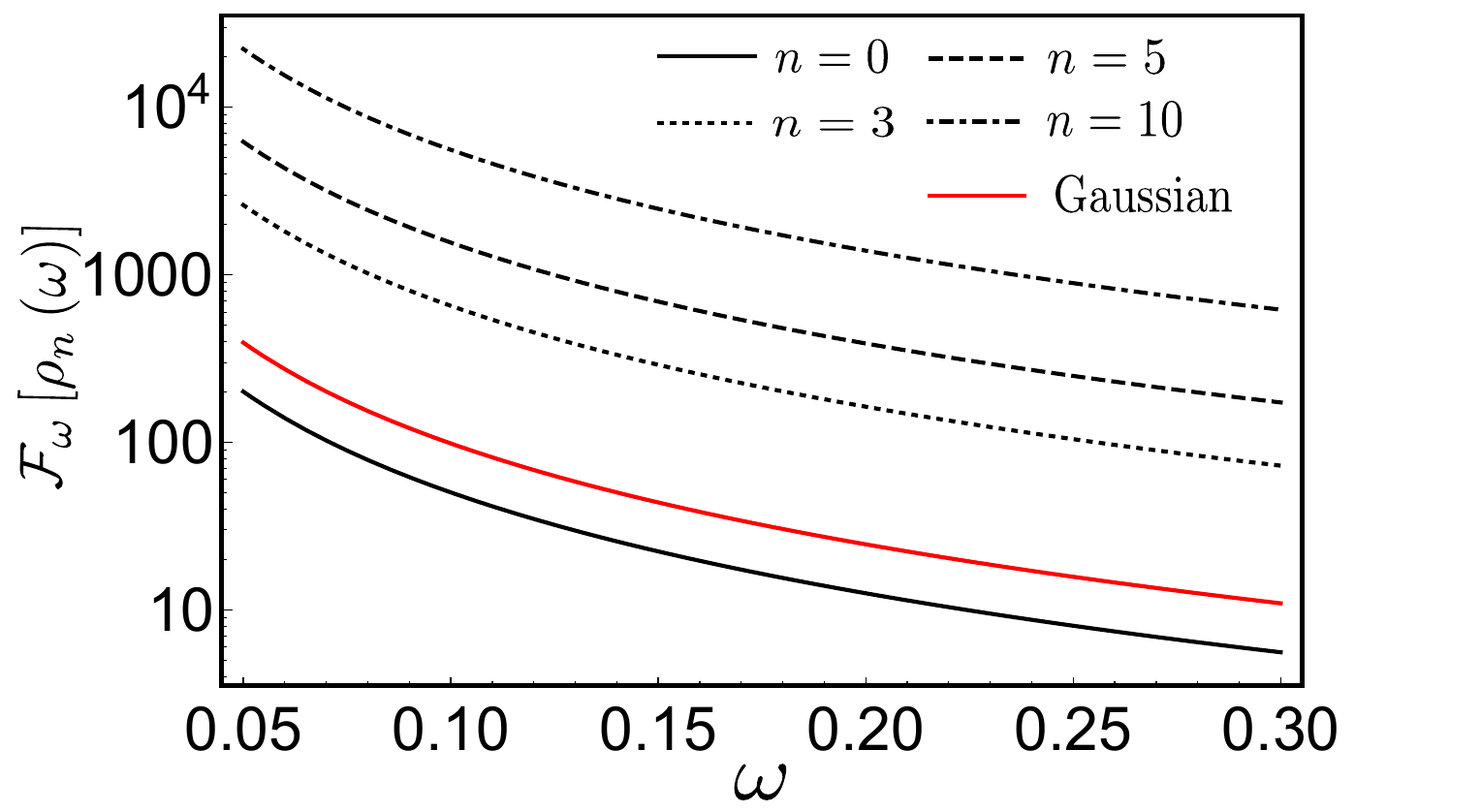}
\caption{Quantum Fisher information $\mathcal{F}_{\omega}\left[\rho_{n}\left(\omega\right)\right]$
as a function of frequency for different Fock states as probe: $n=0$,
vacuum state (black solid line), $n=3$ (black dotted line), $n=5$
(black dashed line), $n=10$ (black dotted-dashed line), and the QFI
evaluated with the Gaussian state $\mathcal{F}_{\omega}\left[\rho_{n}^{G}\left(\vec{d},\boldsymbol{\sigma}\right)\right]$
obtained with the first moments and covariance matrix of the Fock
state $\rho_{n}\left(\omega\right)$ (red solid line).}
\label{QFIfigure}
\end{figure}

From Eq. (\ref{qfi04}) we observe that $\mathcal{F}_{\omega}\left[\rho_{n}^{G}\left(\vec{d},\boldsymbol{\sigma}\right)\right]=\left(2\omega^{2}\right)^{-1}=\mathcal{F}_{\omega}\left[\rho_{n}\left(\omega\right)\right]$
for $n=0$ and rapidly increases to $\omega^{-2}$ as $n$ increases. Generally
speaking, we can say that for $n>0$, $\mathcal{F}_{\omega}\left[\rho_{n}^{G}\left(\vec{d},\boldsymbol{\sigma}\right)\right]\approx\omega^{-2}$
. Figure \ref{QFIfigure} illustrates the QFI for frequency estimation
using different Fock states as probes: $n=0$ (black solid line),
$n=3$ (black dotted line), $n=5$ (black dashed line) and $n=10$
(black dotted-dashed line). In addition, the solid red line represents
the case in which $\mathcal{F}_{\omega}\left[\rho_{n}^{G}\left(\vec{d},\boldsymbol{\sigma}\right)\right]\approx\omega^{-2}$.
From the fact that non-Gaussianity is the only non-classical aspect
concerning the probe states, we can claim that the enhancement in
the QFI is due to the nG-degree of the Fock states, enabling a reduction
in the uncertainty of $\omega$. Although the QFI decreases as $\omega$
increases for every Fock state, the relative advantage between the
Fock state $\rho_{n}$ and the ground state $\rho_{0}$, defined here
as $\epsilon_{n}=10\ln\left\{ \mathcal{F}_{\omega}\left[\rho_{n}\left(\omega\right)\right]/\mathcal{F}_{\omega}\left[\rho_{0}\left(\omega\right)\right]\right\} $
does not depend on the frequency and is $\epsilon_{n}=n^{2}+n+1$,
meaning that the uncertainty $\left(\delta\omega\right)^{2}$ can
be arbitrary reduced provided the experimental ability to generate
higher Fock states.

\subsection{Quantum Fisher information with Fock states and energetic difference}

\begin{figure}
\includegraphics[scale=0.65]{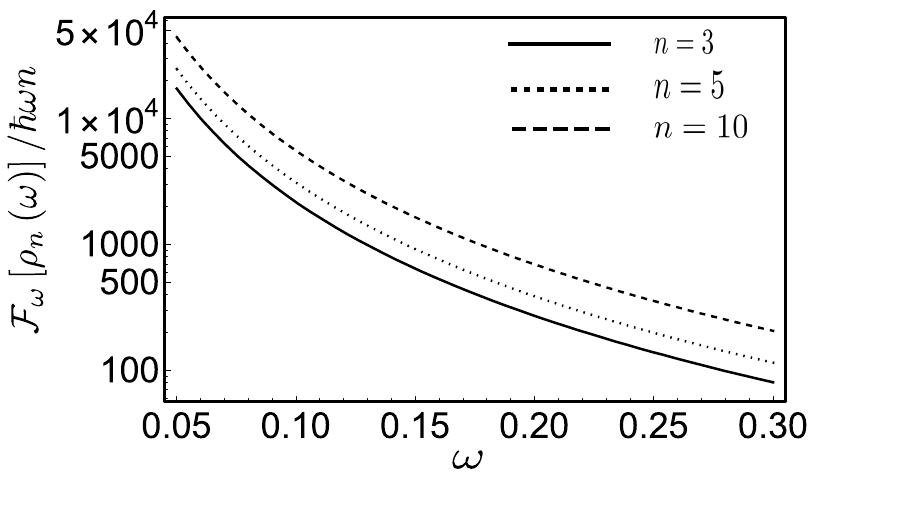}
\caption{Ratio between the quantum Fisher information $\mathcal{F}_{\omega}\left[\rho_{m}\left(\omega\right)\right]$
and the energy difference $\Delta\mathcal{U}=\mathcal{U}\left[\rho_{n}\left(\omega\right)\right]-\mathcal{U}\left[\rho_{0}\left(\omega\right)\right]$
for different Fock states probe. $n=3$ (black solid line), $n=5$
(black dotted line), and $n=10$ (black dashed line).}
\label{qfienergy}
\end{figure}

The enhancement in the frequency estimation due to the nG-degree of
the Fock states naturally depends on the energetic cost to reach these
states. The simplest way to compute this energetic cost is to calculate
the energy difference between the corresponding excited Fock state
and the ground state, $\Delta E_{n}=E\left(\rho_{n}\right)-E\left(\rho_{0}\right)$,
with $E\left(\rho\right)=Tr\left[H\rho\right]$. For the present case,
$\Delta E_{n}=\hbar\omega n$. In Fig. \ref{qfienergy} we show
the ratio $\mathcal{F}_{\omega}\left[\rho_{n}\left(\omega\right)\right]/\hbar\omega n$
for different Fock states, $n=3$ (solid black line), $n=5$ (dotted black line), and $n=10$ (black dashed line). Figure \ref{qfienergy}
evidences that the advantage in exploiting nG-degree of the Fock states
is compatible with the respective energetic cost. The same would not
occur, for example, if $\mathcal{F}_{\omega}\left[\rho_{m}\left(\omega\right)\right]/\hbar\omega m<\mathcal{F}_{\omega}\left[\rho_{n}\left(\omega\right)\right]/\hbar\omega n$
for $m>n$ with a fixed value of $\omega$. In this case, the QFI would not outweighs  the energetic cost of a given excited Fock state.

\subsection{Relation between the quantum Fisher information and the nG-degree}

We have discussed that the only possible non-classical resource of
Fock states is the nG-degree, here quantified in terms of $\delta_{nG}\left[\rho_{n}\right]$.
Although the quantity $\delta_{nG}\left[\rho_{n}\right]$ does not
depend on the frequency $\omega$ (see Fig. \ref{nGfigure}), the
QFI does. This fact motivates the study of how the QFI behaves as
a function of the nG-degree. This point could be relevant for further
experimental proposals, in which both the quantum Fisher information
and the nG-degree could be quantified from a complete measurement
of $\rho_{n}$, i.e., by performing a quantum state tomography. 

\begin{figure}
\includegraphics[scale=0.65]{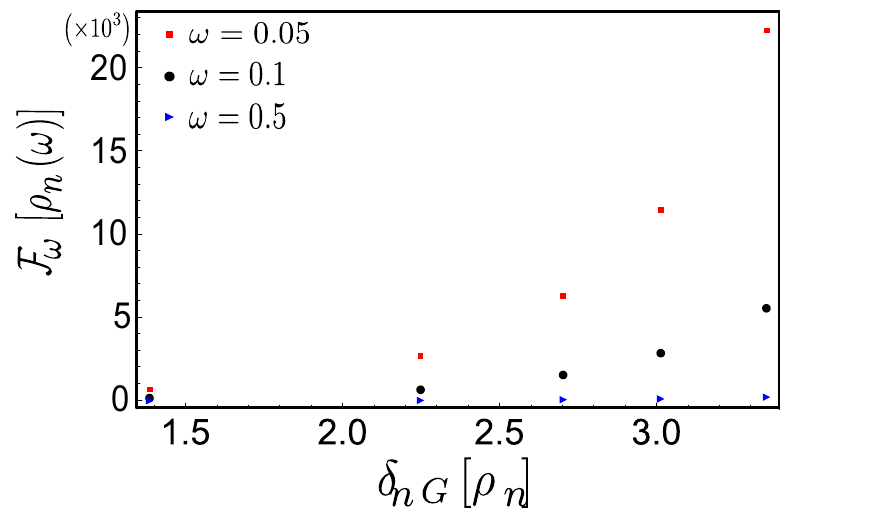}
\caption{Values of the pair $\left(\delta_{nG}\left[\rho_{n}\right],\mathcal{F}_{\omega}\left[\rho_{n}\left(\omega\right)\right]\right)$
for three values of frequencies: $\omega=0.05$ (red squares), $\omega=0.1$
(black circles), and $\omega=0.5$ (blue triangles) and for different
Fock states $n=\left(1,3,5,7,10\right)$, which the ordering corresponds
to the crescent values of $\delta_{nG}\left[\rho_{n}\right]$.}
\label{QFIvsnGFigure}
\end{figure}

We start by considering a given Fock state $\rho_{n}=|\psi_{n}\left(x\right)\rangle\langle\psi_{n}\left(x\right)|$
and compute the pair $\left(\delta_{nG}\left[\rho_{n}\right],\mathcal{F}_{\omega}\left[\rho_{n}\left(\omega\right)\right]\right)$
for a particular value of frequency $\omega$ and different excitations
$n$. Figure \ref{QFIvsnGFigure} depicts the pair $\left(\delta_{nG}\left[\rho_{n}\right],\mathcal{F}_{\omega}\left[\rho_{n}\left(\omega\right)\right]\right)$
for three values of frequencies: $\omega=0.05$ (red squares), $\omega=0.1$
(black circles), and $\omega=0.5$ (blue triangles) and for different
Fock states $n=\left(1,3,5,7,10\right)$. Firstly, we note that for
small nG-degree, $n=1$ and roughly for $n=2$, the QFI varies slightly
for different frequencies. However, as the nG-degree increases, $n=(9,10)$
and larger, the QFI for small frequencies becomes higher. This could
mean that, if we desire to estimate the frequency of a photonic system
for $\omega$ sufficiently small we do not need to access high Fock
states. However, as the value of $\omega$ is increased, one requirement
to reach a larger QFI is the ability to generate higher excited Fock
states. We here refer to Refs.\citep{Cooper2013,Deng2024}, which
experimentally generated excited states with $n=3$ and $n=100$,
respectively.

\section{Generating nG-degree in the photonic probe mode\label{sec:Generating-nG-degree-in}}

\subsection{Transferring non-Gaussianity between two photonic systems}

Despite different mechanisms to generate excited Fock states in a
single cavity \citep{Perarnau-Llobet2020,Deng2024-2}, we can also
consider a photonic system as the probe and its interaction with another
auxiliary system (ancilla) \citep{Santos2021} . The main purpose
here is to present a simple protocol in which non-Gaussianity degree
can be transferred from the ancilla to the probe system, in order
to allow a boost in the frequency estimation. As the resource is assumed
to be the nG-degree of each excited Fock state, we use the Fidelity
between the probe state and some excited Fock state to ensure the
system reached the desired state and then the frequency estimation
can be performed. The step-by-step protocol can be stated as followed:

1. The probe system is initially prepared in the ground state, $\rho_{S}=|0\rangle\langle0|$,
while the ancilla is prepared in some excited Fock state $\rho_{A}=|m\rangle\langle m|$.
The probe and the ancilla are initially uncorrelated, $\rho\left(0\right)=\rho_{S}^{n=0}\left(0\right)\otimes\rho_{A}^{m}\left(0\right)$
.

2. The probe and ancilla interact through the following Hamiltonian

\begin{equation}
H=\omega_{S}a^{\dagger}a+\omega_{A}b^{\dagger}b+\gamma\left(a^{\dagger}b+ab^{\dagger}\right),\label{int01}
\end{equation}
which constitutes a unitary dynamics, with $a\left(a^{\dagger}\right)$
and $b\left(b^{\dagger}\right)$ the bosonic operators of the probe
and ancilla, respectively, and $\gamma$ is the coupling parameter. The interaction Hamiltonian in Eq. (\ref{int01})
means that the two photonic systems will completely exchange their
excitations in an interaction time $t=t_{\ast}$. Furthermore, this interaction is very-well studied and simulated in trapped ions systems \cite{Oliveira2022}. After the dynamics
has started the fidelity between the probe state and the excited Fock
state $|m\rangle\langle m|$ is computed, i.e., $F\left(\rho_{S}^{0}\left(t\right),|1\rangle\langle1|\right)$,
with $F(\rho_{1},\rho_{2})=\left(\text{Tr}\sqrt{\sqrt{\rho_{1}}\rho_{2}\sqrt{\rho_{1}}}\right){}^{2}$.

3. For a given time $t=t_{\ast}$, we will have the unity for the
fidelity, $F\left(\rho_{S}^{0}\left(t\right),|1\rangle\langle1|\right)|_{t=t_{\ast}}=1$.
Then the interaction is finished, and the probe state is obtained
by tracing out the ancilla, $\rho_{S}=Tr_{A}\left[\rho\left(t_{\ast}\right)\right]=|m\rangle\langle m|$.

4. With the nG-degree completely transferred from the ancilla to the
system, the frequency estimation is performed.

Furthermore, it is possible to obtain an expression relating the mutual
information between the system and ancilla with the nG-degree of each
photonic system. The mutual information is given by \citep{Nielsenbook}

\begin{equation}
I_{SA}\left(t\right)=S\left(\rho_{S}\left(t\right)\right)+S\left(\rho_{A}\left(t\right)\right)-S\left(\rho\left(t\right)\right),\label{aa01}
\end{equation}
with $\rho_{S\left(A\right)}\left(t\right)=Tr_{A\left(S\right)}\left[\rho\left(t\right)\right]$.
On the other hand, the nG-degree for each photonic mode is given by
Eq. (\ref{ng01}),

\begin{align}
\delta_{nG,S}\left[\rho_{S}\left(t\right)\right] & =S\left(\rho_{S,G}\left(t\right)\right)-S\left(\rho_{S}\left(t\right)\right),\nonumber \\
\delta_{nG,A}\left[\rho_{A}\left(t\right)\right] & =S\left(\rho_{A,G}\left(t\right)\right)-S\left(\rho_{A}\left(t\right)\right).\label{aa02}
\end{align}

From Eqs. (\ref{aa01}) and (\ref{aa02}) one gets the following relation

\begin{equation}
\Delta I_{SA}\left(t\right)=\Delta S-\delta_{nG,S}\left[\rho_{S}\left(t\right)\right]-\delta_{nG,A}\left[\rho_{A}\left(t\right)\right],\label{003}
\end{equation}
with $\Delta I_{SA}\left(t\right)\equiv I_{SA}\left(t\right)-I_{SA,G}\left(t\right)$
the difference between the mutual information for the state $\rho\left(t\right)$
and for the Gaussian state $\rho_{G}\left(t\right)$, computed with
the first and second moments of $\rho\left(t\right)$, and $\Delta S\equiv S\left(\rho_{G}\left(t\right)\right)-S\left(\rho\left(t\right)\right)$
the difference between the von Neumann entropy using $\rho\left(t\right)$
and $\rho_{G}\left(t\right)$. It is easy to see from Eq. (\ref{003})
that if $\rho_{S}\left(t\right)$ and $\rho_{A}\left(t\right)$ are
both Gaussian, then $\Delta I_{SA}\left(t\right)=0$. However, it
also shows that nG-degree in the local modes can contribute to reduce
the right side in Eq. (\ref{003}), occasionally leading to $\Delta I_{SA}\left(t\right)=0$.

\begin{figure}
\includegraphics[scale=0.5]{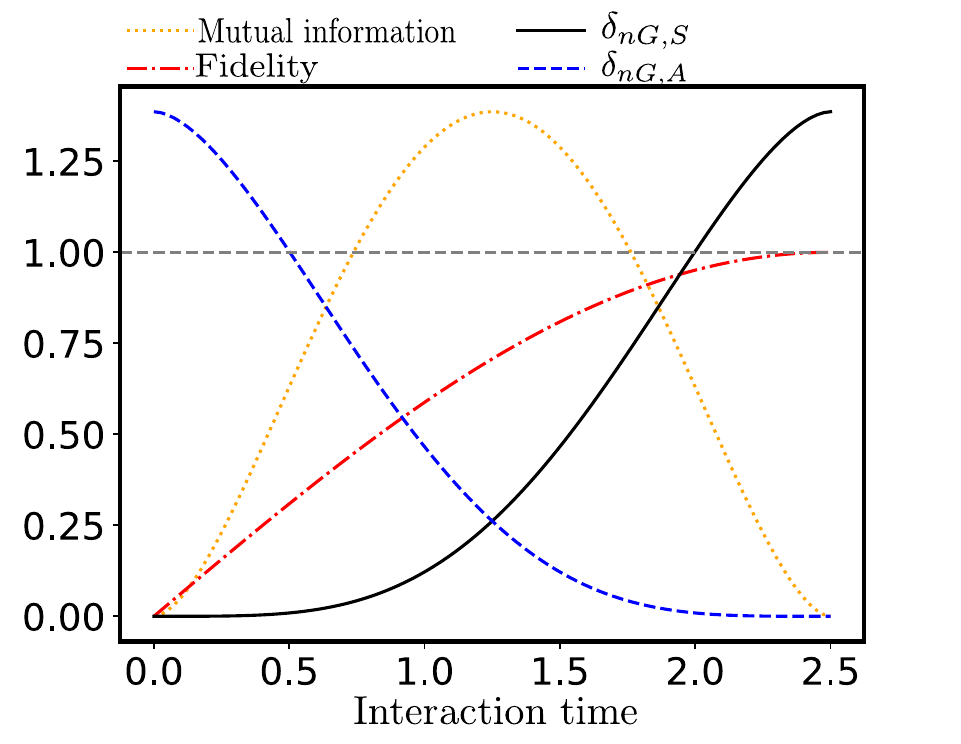}
\caption{Information quantities during the evolution of two photonic modes.
nG-degree of the system (black solid line), nG-degree of the ancilla
(blue dashed line), mutual information between system and ancilla
(orange dotted line), and fidelity between the state of the system
and the Fock state with $n=1$. We have set the coupling parameter to be $\gamma = 0.1$.}
\label{twomodesfigure}
\end{figure}

Figure \ref{twomodesfigure} depicts the information quantities for
two photonic systems evolving accordingly Eq. (\ref{int01}). For
simplicity, we have chosen the initial ancilla state to be $\rho_{A}=|1\rangle\langle1|$.
We can observe the evolution of the nG-degree of the system (black
solid line), nG-degree of the ancilla (blue dashed line), mutual information
between system and ancilla (orange dotted line), and fidelity (red
dashed-dotted line) between the state of the system and the Fock state
with $n=1$. At $t=0$, Fig. \ref{twomodesfigure} shows that the
nG-degree of the system (probe) is zero, because it is prepared in
the ground state. This is also verified by the fidelity. As the interaction
takes place, the mutual information increases, leading to a increase
(reduction) of the nG-degree associated to the system (ancilla). This
is followed by a increase of the fidelity. The dynamics occurs up
to a specific time, which corresponds to a fidelity equal to the unity,
meaning that the system state is in the Fock state $n=1$. At this
moment, the frequency estimation can be performed.

This method can be generalized for the initial state of the ancilla prepared
in other excited Fock state.

\subsection{Photon number measurement on the photonic probe}

We detail another method to generate excited Fock states in a photonic
probe system, by using photon number measurements \citep{Yi2017,Behzadi2021,Santos2023}.
Again, we assume the probe system is initially prepared in the ground
state, $\rho\left(0\right)=|0\rangle\langle0|$. Photon number measurements
acting on a single photonic mode can be described by positive operator-valued
measurements (POVM's). After the measurement protocol, the probe state
is transformed to $\rho^{M}=\sum_{i}M_{i}\rho\left(0\right)M_{i}^{\dagger}$,
with the measurement operators satisfying the condition $\sum_{i}M_{i}^{\dagger}M_{i}=\mathbb{I}$.
In general, photon number measurements can be mediated by a parameter
which controls the measurement intensity, ranging from very weak to
projective measurements. By considering the following measurement
operators $M_{1}=\sum_{n}\sqrt{1-p}|n\rangle\langle n|$ and $M_{2}=\sqrt{p}\sum_{n}|n+1\rangle\langle n|$,
with $0\leq p\leq1$, the state after the measurement is given by

\begin{equation}
\rho^{M}=\left(1-p\right)|0\rangle\langle0|+p|1\rangle\langle1|\label{me01}
\end{equation}

Equation (\ref{me01}) shows the role played by the parameter $p$.
For $p=0$ the measurement do nothing to the state, while for $p=1$
the photonic state is transformed to the first excited Fock state,
and thus the nG-degree could be exploited in the frequency estimation.
Again, here the Fidelity can be used to assure that the probe state
is the first excited Fock state.

\section{QFI for the superposition of Fock states}\label{QFIsuperpositionsection}

In this section, we extend the QFI for a superposition of Fock states for a single photonic mode. The generation of Fock-state superpositions using different protocols has been considered in Ref. \cite{Zhang2024, Shang2004, Yukawa2013} . Consider the following probe state
\begin{equation}
    |\psi\rangle = \frac{1}{\sqrt{N}}\sum_{i=1}^N|i\rangle,
    \label{superposition}
\end{equation}
with $N$ the number of superposition elements. In the position representation we have
\begin{equation}
    \langle x|\psi\rangle =  \frac{1}{\sqrt{N}}\sum_{i=1}^N\langle x|i\rangle,
    \end{equation}
and then $|\partial_\omega \psi \left(x \right) \rangle =\frac{1}{\sqrt{N}}\sum_{i=1}^N |\partial_\omega \psi_i \left(x \right) \rangle$. This implies that
\begin{eqnarray}
    \langle \partial_\omega \psi \left(x \right) | \partial_\omega \psi \left(x \right)\rangle &=& \frac{1}{N} \sum _{i=0}^N \langle \partial_\omega \psi_i \left(x \right)| \partial_\omega \psi_i \left(x \right)\rangle,\nonumber\\
    \langle \psi \left(x \right) | \partial_\omega \psi \left(x \right)\rangle &=& 0
    \end{eqnarray}
where we used the results $\langle \partial_\omega \psi_i \left(x \right)| \partial_\omega \psi_j \left(x \right)\rangle = 0$, $\langle \psi_i \left(x \right) | \partial_\omega \psi_j \left(x \right)\rangle = - \langle \psi_j \left(x \right) | \partial_\omega \psi_i \left(x \right)\rangle$ for $i\neq j$, and $\langle \psi_i \left(x \right) | \partial_\omega \psi_j \left(x \right)\rangle = 0$ for $i=j$.

Then, using Eq. (\ref{qfi02}), the QFI for a superposition of Fock states is
\begin{equation}
    \mathcal{F}_{\textbf{sup}} = \frac{1}{N} \sum_{i=1}^N \mathcal{F}_\omega \left[\rho_i\left(\omega\right)\right].
    \label{supeq}
\end{equation}

\begin{figure}
\includegraphics[scale=0.3]{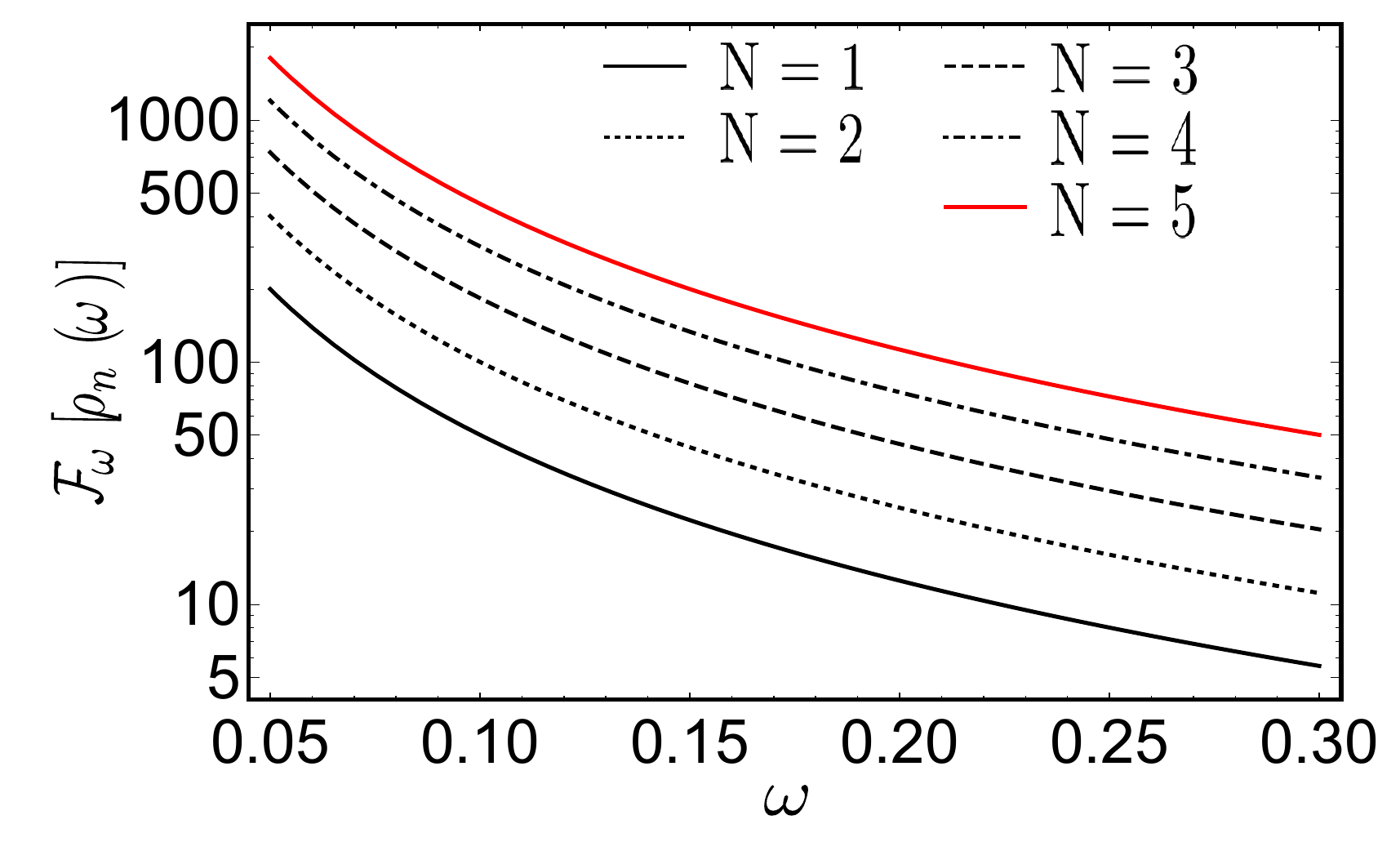}
\caption{Quantum Fisher information for different superposition states as a function of the frequency. $N=1$ ($n = 0$ - black solid line), $N = 2$ ($n = 0$ and $n = 1$ - black dotted line), $N = 3$ ($n = 0$, $n = 1$, and $n = 2$ - black dashed line), $N = 4$ ($n = 0$, $n = 1$, $n = 2$, and $n = 3$ - black dotted-dashed line), and $ = 5$ ($n = 0$, $n = 1$, $n = 2$, $n = 3$, and $n = 4$ - red solid line). }
\label{superfigure}
\end{figure}

Equation (\ref{supeq}) is just a generalization of Eq. (\ref{qfi03}) for the superposition of the Fock states. This results in the same behavior for the QFI, that is, it decreases fundamentally as $\omega^2$. Figure \ref{superfigure} shows the QFI for different superposition states as a function of the frequency: $N=1$ ($n = 0$ - black solid line), $N = 2$ ($n = 0$ and $n = 1$ - black dotted line), $N = 3$ ($n = 0$, $n = 1$, and $n = 2$ - black dashed line), $N = 4$ ($n = 0$, $n = 1$, $n = 2$, and $n = 3$ - black dotted-dashed line), and $ = 5$ ($n = 0$, $n = 1$, $n = 2$, $n = 3$, and $n = 4$ - red solid line). As expected, the greater the number of Fock states in the superposition the higher the QFI for a given value of $\omega$. However, comparing with Fig. \ref{QFIfigure} we observe that the factor $1/N$ makes the QFI worse than the QFI evaluated for a single excited Fock state.

\section{Conclusion\label{sec:Conclusion}}

Recent years have witnessed extensive investigations into intrinsic quantum features to achieve ultimate precision limits in quantum metrology and sensing protocols for relevant physical parameters. This work contributes to this endeavor by exploring the potential enhancement of the quantum Fisher information (QFI) for frequency estimation through the exploitation of non-Gaussianity inherent in the excited states of a photonic mode.

We quantified the degree of non-Gaussianity (nG-degree) based on the relative entropy for photonic excited states and subsequently employed these states as probes for frequency estimation, calculating the corresponding QFI. Our analysis revealed a direct proportionality between the QFI and the nG-degree of the probe states, indicating that for a given excited probe state, lower frequencies yield higher QFI values. Furthermore, we incorporated a straightforward approach to account for the energetic cost associated with the generation of any excited probe state. The ratio of the QFI to the energy difference demonstrated that the gain in QFI outweighs the energetic expenditure. We also briefly discussed two potential avenues for generating non-Gaussianity in photonic probes: interaction with an ancillary bosonic mode and photon number measurement with high parametric control.

Several open questions warrant further investigation. Exploring the role of non-Gaussianity in quantum metrology protocols within systems beyond photonic ones presents a compelling direction. Notably, the gravitational quantum well, where eigenstates are described by Airy functions, constitutes an interesting system for such studies in gravimetry. Additionally, the synergistic utilization of non-Gaussianity as an auxiliary resource in metrological protocols that leverage other quantum resources, such as coherence, remains unexplored. To summarize, the observed relationship between the QFI for frequency estimation and the nG-degree is, in principle, experimentally verifiable through the reconstruction of the Wigner function via quantum state tomography and the subsequent analysis of statistical moments. We anticipate that our findings can contribute to elucidating the role of non-Gaussianity in quantum information protocols and, in particular, motivate further experimental investigations.

\begin{acknowledgments}
Jonas F. G. Santos acknowledges CNPq Grant No. 420549/2023-4, Fundect
Grant No. 83/026.973/2024, and Universidade Federal da Grande Dourados
for support.
\end{acknowledgments}

\end{document}